\documentclass[doublecol]{epl2}

\title{Creep motion of a domain wall in the two-dimensional random-field Ising model with a driving field}

\author{R. H. Dong, B. Zheng\footnote{corresponding author; email: zheng@zimp.zju.edu.cn} and N. J. Zhou}
\shortauthor{R. H. Dong$^{1}$, B. Zheng$^{1}$ and N. J. Zhou$^{2}$}

\institute{
$^1$D epartment of Physics, Zhejiang University, Hangzhou 310027,
P.R. China\\
$^2$ Department of Physics, Hangzhou Normal University, Hangzhou 310036, China}


\abstract{With Monte Carlo simulations, we study the creep motion of a domain wall in the two-dimensional
random-field Ising model with a driving field. We observe the nonlinear field-velocity relation,
and determine the creep exponent $\mu$. To further investigate the universality class of the creep motion,
we also measure the roughness exponent $\zeta$ and energy barrier exponent $\psi$ from the zero-field relaxation process.
For strong disorder, the exponents are consistent with those of the Edwards-Wilkinson equation;
for weak disorder, a different universality class is detected.}

\begin{document}

\maketitle

\section{Introduction}

In recent years the dynamics of elastic systems in disordered media has been a focus of theoretical and experimental studies.
Examples are charge-density waves, vortex lattices, domain walls in ferromagnetic or ferroelectric materials, contact lines,
and fluid invasion in porous media \cite{he92,lem98,ono99,mou04,yam07,im09}. In particular, the magnetic domain-wall dynamics
is an important topic in magnetic devices, nanomaterials, thin films, and semiconductors \cite{met07,shi07,dou08,kim09}.
At zero temperature, the domain-wall motion exhibits a depinning transition. Due to the energy barriers created by the disorder,
the domain wall is pinned and the velocity of the domain wall remains zero up to a critical field $H_{c}$ \cite{now98,due05,kol06a,bak08}.
At finite temperature, the depinning transition is softened and the energy barriers can always be overcome by thermal activation.
Thus, with a sufficiently small field, $H \ll H_{c}$, the domain-wall motion can reach a steady state,
which is known as the creep motion \cite{lem98,kol05a,rot01,cha00}.

Up to date, most theoretical approaches to the domain-wall dynamics in ferromagnetic materials
are typically based on the Edwards-Wilkinson equation with quenched disorder (QEW). This equation is a phenomenological model,
and detailed microscopic structures and interactions of real materials are not concerned.
To further understand the domain-wall motion from a more fundamental level, we should build lattice models
based on microscopic structures and interactions.
The random-field Ising model with a driving field (DRFIM) is a candidate,
at least to capture robust features of the domain-wall motion,
although it does not include all interactions in real materials.
It may go beyond the QEW equation, reveal new universality classes, explain a wider range of physical phenomena,
and allow a closer comparison with experiments.
For example, the depinning transition of the two-dimensional DRFIM model has been examined
with the short-time dynamic approach \cite{zho09,zho10}.
The critical exponents are accurately determined, and the results indicate that the DRFIM model and QEW equation
are not in a same universality class.
In a very recent work, it is demonstrated that the relaxation state and
relaxation-to-creep transition of the domain-wall motion can be simulated
with the DRFIM model, and the Cole-Cole plot of the complex susceptibility obtained in the experiments is nicely reproduced \cite{zho10a,kle07}.

The main purpose of this paper is to study the creep motion of the domain wall in the two-dimensional DRFIM model, in comparison with
the QEW equation and experiments. In the creep regime of the domain-wall motion,
the field-velocity relation is expected to be
\begin{equation}
v \sim exp[-(H_{c}/H)^{\mu}/T], \label{vHrel}
\end{equation}
where $\mu$ is the creep exponent, usually used to classify the universality class of the materials
in experiments \cite{lem98,yam07,lee11}.
Based on the QEW equation,
two universality classes have been identified for the field-induced domain-wall motion
in two dimensions \cite{cha00,gia06}: $\mu=1/4$ belongs to the random-bond disorder, whereas $\mu=1$ to the random-field disorder, respectively.
In the past years, most experiments confirm this result.
However, recent progress in experiments suggests that the degree of surface roughness may alter the universality class \cite{kan10}.
Different values of $\mu$ between $1$ and $1/4$ are also reported \cite{par05,per08}.
Due to the uncertainty of the domain-wall dimensionality, it leads to conflicts in identifying the universality classes
of the materials \cite{par05,jo09}.
These experimental results challenge our theoretical understanding. In the QEW equation,
the degree of disorder is thought to be irrelevant to the universality class.

On the other hand, based on the QEW equation, one may theoretically derive the scaling law \cite{kol05a,kol09}
\begin{equation}
\mu=\frac{2\zeta+d-2}{2-\zeta}, \label{crprel}
\end{equation}
where $\zeta$ is the equilibrium roughness exponent.
In the numerical study of the QEW equation, however, a violation of the scaling law in Eq.~(\ref{crprel})
is observed for low temperatures or with strong disorder \cite{kol05a},
which implies a more complicated physical scenario. Moreover, it is argued in Ref.~\cite{mon08} that instead of Eq.~(\ref{crprel}),
the creep exponent should obey the scaling law
\begin{equation}
\mu=\frac{\psi}{2-\zeta}, \label{mpzrel}
\end{equation}
where $\psi$ is the energy barrier exponent without a driving field \cite{mon08,kol09a}.
The scaling laws in Eqs.~(\ref{crprel}) and (\ref{mpzrel}) will be equivalent
if $\psi$ is identical to the ground-state-energy fluctuation exponent $\theta$,
\begin{equation}
\psi=\theta \equiv 2\zeta+d-2. \label{pzrel}
\end{equation}
Theoretically, there are still some controversies on the value of $\psi$ \cite{kol05a,mon08,noh09}.

In order to fully understand the creep motion and its universality class, it is important to directly determine the exponents $\psi$ and $\zeta$.
This can be done by independently simulating the zero-field relaxation process of a domain wall \cite{kol05a,noh09}.
Since the driving field is absent, the correlated domain-wall segment of a length $\xi$
overcomes the energy barrier $U(\xi)$ only by thermal activation. According to the Arrhenius law,
the time scale of this activation process is the order of $exp[U(\xi)/T]$.
It is believed that the energy barrier may scale as $U(\xi) \sim \xi^{\psi}$ for a large $\xi$.
Therefor, at long times, the correlation length should grows by a logarithmic law
\begin{equation}
\xi(t) \sim [T\ln (t/t_0)]^{1/\psi}. \label{xprel}
\end{equation}
As the domain wall propagates, it continues to roughen as an interface.
The roughness function of the domain interface is believed to scale as \cite{yan95,jos96,zho08,bak08,zho09}
\begin{equation}
\omega^{2}(t) \sim 2\xi(t)^{\zeta}, \label{sclw2}
\end{equation}
where $\zeta$ is the equilibrium roughness exponent.

Early numerical study of the DRFIM model also intended to tackle the creep motion \cite{rot01}.
However, due to the limited range of the driving fields, the validity of Eq.~(\ref{vHrel}) is not conclusively confirmed.
Further understanding of the scaling laws in Eqs.~(\ref{mpzrel}) and (\ref{pzrel}) is also lacking.
In particular and more importantly, in this paper, we investigate the possible dependence of the universality class
on the strength of the disorder.

\section{The model}

The two-dimensional DRFIM model is defined by the Hamiltonian
\begin{equation}
\mathcal{H} = - J \sum_{<ij>}S_iS_j - \sum_i (h_i+H) S_i,
\label{defH}
\end{equation}
where $S_i = \pm 1$ is an Ising spin on the square
lattice. The quenched random field $h_i$ is uniformly distributed within an
interval $[-\Delta, \Delta]$, and $H$ is a homogeneous driving field.
In this paper, we take the coupling constant $J=1$. Since the two-dimensional random-field Ising model
does not have a long-range ordered phase, one expects that domains grow spontaneously in the bulk at a finite temperature.
This may lead to ambiguity in defining a single domain interface. If it is at low enough temperatures,
however, the dynamic evolution of the bulk is negligible. In our study, the temperature is restricted to $T_{max}=0.67$.

Our simulations are performed on a lattice, with a linear size $2L$ in the $x$ direction and $L$ in the $y$ direction.
Antiperiodic and periodic boundary conditions are adopted in the $x$ and $y$ directions, respectively.
To eliminate the pinning effect irrelevant for the quenched disorder,
we rotate the square lattice such that the initial domain wall orients
in the $(11)$ direction of the square lattice, as shown in Refs.~\cite{now98,rot99,rot01,zho09}.

The initial state is a {\it semiordered} state with a perfect domain wall in the $y$ direction.
After preparing the initial state, we {\it randomly} select a spin, and flip it if the total energy decreases after flipping.
A Monte Carlo time step is defined by $2L^2$ single-spin flips. Simulations are performed at two temperatures $T=0.33$ and $0.67$
with lattice size $L=256$ up to $t_{max}=10^6$. For the strength of random fields, two typical values of $\Delta$ are chosen,
i.e., $\Delta = 0.5$ and $1.2$.
Simulations of different $L$ are also performed to confirm that the finite-size effects are already negligibly small.
For the simulations of the creep motion, the domain-wall velocity is averaged by at least $1000$ samples.
For the zero-field relaxation process, the total samples are $5000$.
Errors are estimated by dividing the samples into three or four subgroups.

To study the creep motion and zero-field relaxation of the domain interface, we first introduce a {\it line} magnetization
\begin{equation}
m(y,t) = \frac{1}{L} \left[ \sum_{x=1}^L S_{xy}(t) \right].
 \label{defm}
\end{equation}
Here $S_{xy}(t)$ denotes a spin at site $(x, y)$.
The height function of the interface is defined as
\begin{equation}
h(y,t) = \frac{L}{2}[ m(y,t) + 1] . \label{defh}
\end{equation}
With the height function at hand, the average velocity of the interface can be calculated
\begin{equation}
v(t) = \left \langle \frac{dh(y,t)}{dt}\right \rangle, \label{defv}
\end{equation}
where $<\cdots>$ includes the statistical average and average over
$y$.

With the height function $h(y,t)$, the roughness function of the interface is defined as
\begin{equation}
\omega^{2}(t) = \left \langle h(y,t)^2 \right \rangle -
\langle h(y,t) \rangle^2. \label{defw2}
\end{equation}
A more informative quantity is the height correlation function
\cite{jos96},
\begin{equation}
C(r, t) = \left\langle[h(y + r, t) - h(y, t)]^2 \right
\rangle. \label{defC}
\end{equation}
It describes both the spatial correlation of the height function in
the $y$ direction and the growth of the domain interface in the $x$
direction.

For a sufficiently large lattice, i.e., the correlation length $\xi(t) \ll L$, we should
observe the standard scaling behavior for the roughness function in Eq.~(\ref{sclw2}),
and that for the height correlation function \cite{yan95,jos96,zho08,bak08,zho09},
\begin{equation}
   C(r,t)=2\omega^2(t)f(r/\xi). \label{sclC1}
\end{equation}

\section{Monte Carlo simulations}

To investigate the characteristic relation between the driving field $H$ and domain-wall velocity $v$,
we measure the velocity for a given driving field when the steady state is reached.
Theoretically the creep regime is expected to be far below the critical point $H_{c}$.
For a smaller driving field $H$, however, it takes a longer time to reach the steady state.
In our simulations, the longest time is $10^6$ Monte Carlo time steps.

\begin{figure}[t]
\includegraphics[scale=0.3]{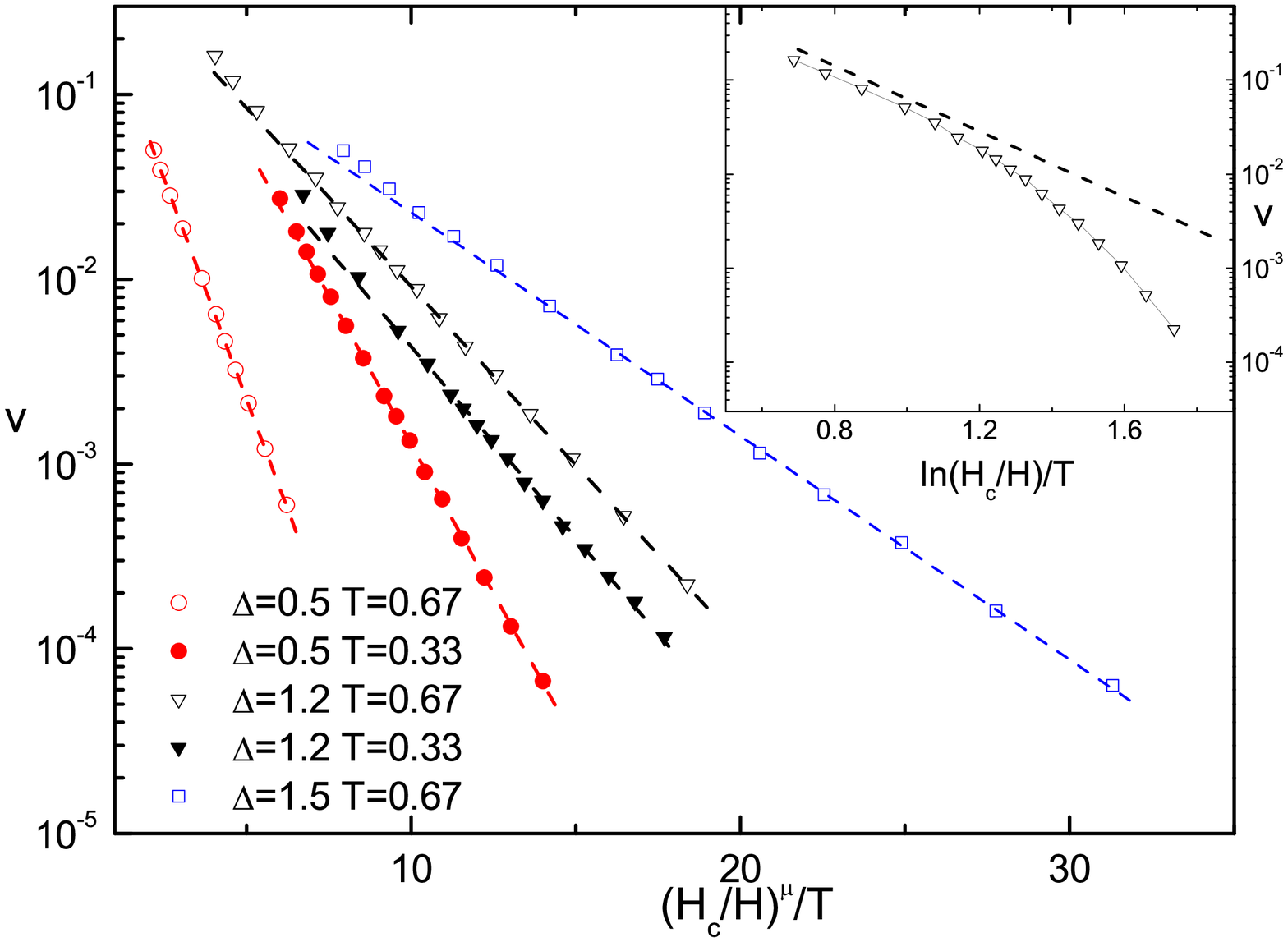}
\caption{The domain-wall velocity $v$ is plotted as a function of $(H_{c}/H)^{\mu}/T$
for different disorders and temperatures on a semi-log scale.
Dash lines show an exponential behavior.
For $\Delta=1.2$, $\mu=1.02(5)$ and $0.95(8)$ at $T=0.67$ and $0.33$ respectively;
for $\Delta=0.5$, $\mu=0.63(5)$ and $0.59(4)$ at $T=0.67$ and $0.33$ respectively.
In the inset, $v$ is plotted against $\ln (H_{c}/H)/T$ for $\Delta=1.2$ and $T=0.67$ on a semi-log scale.
Errors of the data points are smaller than the symbol sizes.}\label{f1}
\end{figure}

In Fig.~\ref{f1}, we plot the velocity $v$ as a function of $(H_{c}/H)^{\mu}/T$.
By adjusting the exponent $\mu$, one may observe a linear curve. Thus the nonlinear field-velocity relation
in Eq.~(\ref{vHrel}) is verified in the two-dimensional DRFIM model.
For $\Delta=1.2$, we obtain $\mu=1.02(5)$ and $0.95(8)$ at $T=0.67$ and $0.33$ respectively,
in good agreement with the theoretical prediction $\mu=1$ based on the QEW equation.
The small deviation from the exponential law for large fields suggests that the system leaves the creep regime.
For $\Delta=0.5$, however, the fitted exponent is $\mu=0.63(5)$ and $0.59(4)$ at $T=0.67$ and $0.33$ respectively,
significantly smaller than the theoretical value $\mu=1$ of the QEW equation.
To further confirm our results, we may plot $\ln v$ against $H_{c}/H$ on a log-log scale.
From slopes of the curves, we measure the exponent $\mu$. Or, we may directly  fit the numerical data to the
formula in Eq.~(\ref{vHrel}) to extract $\mu$. All these techniques yield similar values for the exponent $\mu$.

In Ref.~\cite{rot01}, it is stated that the numerical data could fit to the field-velocity relation
in Eq.~(\ref{vHrel}) in the limit $\mu \to 0$.
In the inset of Fig.~\ref{f1}, we plot $v$ against $ln(H_{c}/H)/T$ on a semi-log scale.
A tendency of a linear behavior is observed for large driving fields, supporting the results in Ref.~\cite{rot01}.
Therefore we believe that the field regime considered in Ref.~\cite{rot01} is too large to observe the creep motion.

\begin{figure}[t]
\includegraphics[scale=0.3]{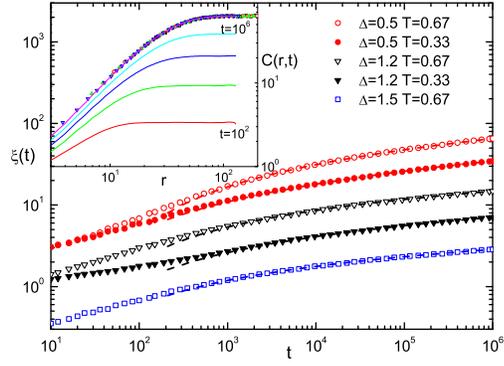}
\caption{The correlation length $\xi(t)$ is displayed for different disorders and temperatures on a log-log scale.
The curves of $\Delta=1.2$ are shifted for clarity. Dashed lines represent logarithmic fits to Eq.~(\ref{xprel}).
For $\Delta=1.2$, $\psi=1.05(7)$ and $0.98(9)$ at $T=0.67$ and $0.33$ respectively.
For $\Delta=0.5$, $\psi=0.69(5)$ and $0.65(6)$ at $T=0.67$ and $0.33$ respectively.
In the inset, the height correlation function $C(r,t)$ of $\Delta=1.2$ and $T=0.67$ at different times
are displayed with solid lines. Symbols show the data collapse according to Eq.~(\ref{sclC1}.)  }\label{f2}
\end{figure}

To verify the scaling laws in Eqs.~(\ref{mpzrel}) and (\ref{pzrel}),
one needs to determine the exponents $\psi$ and $\zeta$ independently.
Thus we simulate the zero-field relaxation process of the domain wall.
In the absence of the driving field $H$, the growth law of the correlation length $\xi(t)$ plays a crucial role.
To extract the correlation length $\xi(t)$, we use a data collapse technique
based on the scaling behavior of the height correlation function.
According to Eq.~(\ref{sclC1}), we fix $t'=10^6$,
and rescale $r$ of another $t$ to $\xi(t')/\xi(t)r$ and $C(r,t)$ to $\omega(t')/\omega(t)C(r,t)$.
If the ratio of $\xi(t')/\xi(t)$ is properly chosen, $C(r,t)$ at different time $t$ collapse onto a single curve.
This is shown in the inset of Fig.~\ref{f2}.
In other words, one may estimate the ratio $\xi(t')/\xi(t)$ from the data collapse of $C(r,t)$.
Thus, up to a constant $\xi(t_0)$, the growth law $\xi(t)$ is extracted.
Alternatively, one may also determine $\xi(t)$ by fitting $C(r,t)$ to the scaling function \cite{jos96}
\begin{equation}
C(r,t)=A[tanh(r/B)]^{2\alpha}, \label{sclC3}
\end{equation}
where  $B$ is interpreted as the correlation length at $t$. Both methods yield the same results.

In Fig.~\ref{f2}, $\xi(t)$ is shown as a function of $t$ on a log-log plot. A crossover behavior is observed.
The correlation length seems to be described by a power law at early times, then obviously slows down after the transient regime,
which implies a logarithmic growth. To extract the energy barrier exponent $\psi$,
we fit the correlation length to Eq.~(\ref{xprel}) in the time regime $t_{0}<t<10^{6}$.
By increasing $t_0$ until the fitted parameters become stabilized,
we observe the logarithmic growth law for almost three decades of time, shown by dash lines in Fig.~(\ref{f2}).
The exponent $\psi \approx 1$ for $\Delta=1.2$ is consistent with the theoretical prediction based on the QEW equation.
In contrast, we obtain $\psi\approx 0.7$ for $\Delta=0.5$.

\begin{figure}[t]
\includegraphics[scale=0.3]{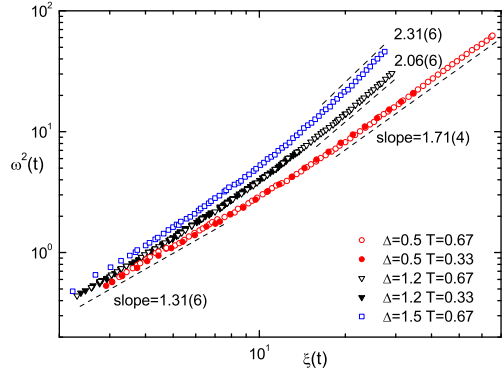}
\caption{The roughness function $\omega^2(t)$ is plotted against $\xi(t)$ on a log-log scale.}\label{f3}
\end{figure}

With the correlation length $\xi(t)$ at hand, one may measure the roughness exponent $\zeta$ from Eq.~(\ref{sclw2}).
In Fig.~\ref{f3}, $\omega^2(t)$ is plotted against $\xi(t)$ on a log-log scale. A crossover behavior is observed.
In the large-$\xi(t)$ regime, the slopes of the curves yields
$\zeta=1.03(3)$ and $0.85(2)$ for $\delta=1.2$ and $0.5$, respectively.
Here we remind ourself that $\xi(t)$ grows very slowly by a logarithmic law, therefore it remains less than $10^2$
even when $t$ reaches $10^6$ Monte Carlo time steps.

In Table~\ref{t1}, all the measurements of the exponents $\mu$, $\psi$, $\zeta$, $\theta$, and $\psi/(2-\zeta)$ are listed.
For comparison, the QEW equation predicts that all these exponents are equal to 1 \cite{fis86,cha00}.
For strong disorder, i.e., $\Delta=1.2$, the exponents are consistent
with those of the QEW equation. For weak disorder, i.e., $\Delta=0.5$, however, the exponents are obviously different,
indicating a different universality class. Especially, the exponent $\mu \approx 0.6$ is between
the random-bold universality class $\mu =1/4$ and the random-field universality class $\mu =1$
predicted by the QEW equation. This may give hints for understanding the experiments which yield
a similar exponent $\mu$ \cite{par05,per08,jo09,kan10}.
On the other hand, although the exponents for the weak disorder depart from those for the strong disorder,
the scaling laws in Eqs.~(\ref{mpzrel}) and (\ref{pzrel}) always hold, i.e., $\psi \approx \theta$
and $\mu \approx \psi/(2-\zeta)$ within statistical errors.

Why do the exponents depend on the strength of the disorder?
At zero temperature, for $\Delta < 1$, the depinning phase transition is of first order,
and the transition field $H_c=\Delta$, since no overhangs can be created,
while for $\Delta > 1$, the depinning phase transition is of second order,
and the transition field $H_c < \Delta$, since overhangs can be generated \cite{ji91,now98,zho09,zho10}.
For the creep motion at nonzero temperatures, the driving field $H$ is much smaller
than $\Delta$ and $H_c$, and the thermal activation plays an important role
in driving the domain wall to propagate. For $\Delta < 1$, however, overhangs can hardly be
generated at low temperatures. In other words, a stronger disorder may induce overhangs,
and overhangs may change the degree of the interface roughness, and alter the universality class of
the domain interface.

In the small-$\xi(t)$ regime, the roughness exponent is apparently smaller than that in the large-$\xi(t)$ regime,
for example, $\zeta=0.65(3)$ for $\Delta=0.5$, which is close to $\zeta=2/3$ of the random-bond disorder \cite{hus85,kar85}.
Interestingly, as shown in Fig.~\ref{f3}, the curves of different temperatures
with a same strength of the disorder collapse onto a single curve,
which implies that the temperature only affects the growth law of the correlation length $\xi(t)$.
This suggests that there may exist a temperature-independent characteristic length $L_c$ in the dynamic relaxation process.
When the correlation length $\xi(t)$ is smaller than $L_c$, $\xi(t)$ grows by a power law and the roughness exponent $\zeta=0.65(3)$,
which may be comparable with the random-bond universality. When the correlation length $\xi(t)$ is larger than $L_c$,
the growth law of $\xi(t)$ crosses over to a logarithmic one.

\section{Conclusion}

To summarize, we have simulated the creep motion of a domain wall in the two-dimensional
random-field Ising model with a driving field. The nonlinear field-velocity relation in Eq.~(\ref{vHrel}) is confirmed,
and the creep exponent $\mu$ is determined. Meanwhile,
we also measure the roughness exponent $\zeta$ and energy barrier exponent $\psi$ from the zero-field relaxation process.
For strong disorder, all the exponents are consistent with those predicted by the Edwards-Wilkinson equation;
for weak disorder, however, the exponents are very much different, indicating a different universality class.
Although the exponents change with the strength of the disorder,
the scaling laws in Eqs.~(\ref{mpzrel}) and (\ref{pzrel}), and therefore also in Eq.~(\ref{crprel}), always hold.

\acknowledgments This work was supported in part by NNSF of China
under Grant Nos. 10875102 and 11075137, and Zhejiang Provincial
Natural Science Foundation of China under Grant No. Z6090130.


\begin{table}
 \centering \footnotesize
\begin{tabular}[t]{c | c c| c c}
\hline
    & \multicolumn{2}{c|}{ $\Delta=1.2$ }  & \multicolumn{2}{c}{ $\Delta=0.5$ } \\
\hline
 T &\quad$0.67$\quad   &   \quad$0.33$\quad  &   \quad$0.67$\quad   &    \quad$0.33$\quad  \\
\hline
$\mu$\quad                       &     \quad1.02(5)   &  \quad0.95(8)    &  \quad0.63(5)  &  \quad0.59(4)           \\
$\psi$\quad                       &      \quad1.05(7)       &     \quad0.98(9)   & \quad0.69(5)   &  \quad0.65(6)        \\
\hline
$\zeta$\quad                &       \multicolumn{2}{c|}{1.03(3)}   &  \multicolumn{2}{c}{ 0.85(2) }\\
$\theta$\quad                              &       \multicolumn{2}{c|}{ 1.06(6) }    &    \multicolumn{2}{c}{ 0.71(4) }    \\
\hline
$\psi/(2-\zeta)$\quad        &     1.08(8)            &  1.01(9)      &  0.60(5)   &    0.57(6)     \\
\hline
\end{tabular}
\caption{Exponents for the two-dimensional DRFIM model.} \label{t1}
\end{table}

\end{document}